\documentclass[12pt, prd, showpacs]{revtex4}
\usepackage{amssymb}
\usepackage{amsmath}
\usepackage{graphicx}

\begin{document}

\title{Inverted equation of state and general approach to vacuum-like
configurations}
\author{O. B. Zaslavskii}
\affiliation{Department of Physics and Technology, Kharkov V.N. Karazin National
University, 4 Svoboda Square, Kharkov 61022, Ukraine}
\email{zaslav@ukr.net }

\begin{abstract}
We consider spherically symmetric static black hole configurations that obey
the vacuum equation of state: $p_{r}=-\rho $, where $p_{r}$ is the radial
pressure, $\rho $ being energy density. We find in a closed form the metric
for an arbitrary equation of state for tangential pressure $p_{\theta }(\rho
)$. The corresponding formulas enable us to embrace compact
Schwarzschild-like configurations and dispersed systems. They include
metrics with a regular center and singular ones. In a particular case, the
metric of the Kiselev black hole is reproduced.
\end{abstract}

\maketitle

\section{Introduction}

Typically, black holes contain singularities hidden under the horizon and
one is led to special efforts to cure them. For spherically symmetric
configurations (which is our subject) this is the point (or, more precisely,
hypersurface) where the coordinate $r=0$. There are different ways to solve
this task. This is a rather vast subject and we refer the reader to a review 
\cite{ans} and references therein ((including, for example, black universes,
models in nonlinear electrodynamics, etc.). In the present Letter, we
concentrate only on one but important class of such objects - the so-called
vacuum-like configurations with the equation of state $p_{r}=-\rho $ where $%
p_{r}$ is the radial pressure and $\rho $ is the energy density \cite{glin}, 
\cite{dym}. This allows us to have a regular center since near $r=0$ the
geometry behaves similarly to the de Sitter metric \cite{dym}.

Recently, a new wave of interest has arisen to constructing models of
regular black holes. They can be considered as an alternative to so-called
black bounces \cite{bron}. Also, in refs. \cite{ov1} - \cite{ov-k2} it was
studied how one can achieve compact configurations with a regular center
that look like the Schwarzschild black hole for an external observer.
Actually, the metrics discussed there also belong to the aforementioned
vacuum-like class.

Meanwhile, neither in the aforementioned works nor in \ the pioneering work 
\cite{dym} a concrete, physically relevant equation of state that relates
the tangential pressure to the energy density $p_{\theta }(\rho )$ was
suggested. Instead, some trial configurations with dependence $\rho (r)$ (or
the mass corresponding to such a density) were taken by hand. Another method
to obtain regular black holes consists in essentially modifying of the
metric near the center \cite{bard}, \cite{vis}.

In addition to regular vacuum-like black holes, their singular counterparts
are also of interest. For example, for the linear equation of state
configurations describing a black hole surrounded by quintessence have been
found \cite{kis}. See also their recent modification in \cite{sant}.

We suggest another approach. We consider vacuum-like configurations with an
arbitrary (in general, non-linear) given equation of state $p_{\theta }(\rho
)$. The main point consists of the interchange of the roles of independent
variable and the unknown function. We take advantage of the fact that a
static spherically symmetric vacuum-like configuration admits closed
formulas for the radial coordinate $r$ as a function of $\rho $. This is a
key observation for what follows below. In other words, instead of
dependence $\rho (r)$ we deal with the dependence $r(\rho )$. In some
physically relevant cases the final formulas can be inverted and give the
metric as a function of $r$. In particular, the exact solutions \cite{dym}, 
\cite{kis} are recovered.

Below, we use geometric units in which the fundamental constants $G=c=1$.

\section{General approach}

We consider a spherically symmetric metric sourced by the stress-energy
tensor%
\begin{equation}
T_{\mu }^{\nu }=diag(-\rho \text{, }p_{r}\text{, }p_{\theta }\text{, }%
p_{\theta })\text{.}
\end{equation}

Assuming that%
\begin{equation}
p_{r}=-\rho \text{,}  \label{eos}
\end{equation}%
one can infer from the Einstein equations that

\begin{equation}
ds^{2}=-Vdt^{2}+\frac{dr^{2}}{V}+r^{2}d\omega ^{2}.  \label{met}
\end{equation}

Let us introduce the mass function $m(r)$ as usual:%
\begin{equation}
V=1-\frac{2m(r)}{r}.  \label{V}
\end{equation}%
It follows from the Einstein equations that%
\begin{equation}
p_{\theta }=-\frac{m^{\prime \prime }}{8\pi r},  \label{m''}
\end{equation}%
\begin{equation}
\rho =\frac{m^{\prime }}{4\pi r^{2}},  \label{m'}
\end{equation}%
where prime denotes derivative of a function with respect to its argument.

Assuming that $m(0)=0$, one infers from (\ref{m'}) that%
\begin{equation}
m(r)=4\pi \int_{0}^{r}dr^{\prime }\rho (r^{\prime })r^{\prime 2}.  \label{m}
\end{equation}

With such a constant of integration, the center is regular, so integration
is taken from $r=0$.

Then, it follows from (\ref{m''}) and (\ref{m'}) that%
\begin{equation}
r=const\exp (-\frac{1}{2}\int^{\rho }\frac{d\rho ^{\prime }}{f(\rho ^{\prime
})})\text{,}  \label{ro}
\end{equation}%
where%
\begin{equation}
f(\rho )\equiv p_{\theta }+\rho \text{.}
\end{equation}

Depending on a type of the configuration we are looking for, we must choose
the property of the function $f(\rho )$ and a constant of integration in (%
\ref{ro}) accordingly.

\section{Regular center}

One can see that for the metric (\ref{met}) under discussion, it is
sufficient to have finite $\rho $ at $r=0$ under the assumption $m(0)=0$ for
the center to be regular. Indeed, according to eq. (14.42) of \cite{mtw},
the nonzero Riemann tensor tetrad components are given by 
\begin{equation}
E\equiv R_{\hat{t}\hat{r}}^{\hat{t}\hat{r}}=-\frac{V^{\prime \prime }}{2},
\label{e}
\end{equation}%
\begin{equation}
\bar{E}\equiv R_{\hat{t}\hat{\theta}}^{\hat{t}\hat{\theta}}=-\frac{V^{\prime
}}{2r},  \label{e1}
\end{equation}%
\begin{equation}
H\equiv R_{\hat{\phi}\hat{\theta}}^{\hat{\phi}\hat{\theta}}=\frac{1-V}{r^{2}}%
,  \label{f}
\end{equation}%
\begin{equation}
\bar{H}\equiv R_{\hat{\theta}\hat{r}}^{\hat{\theta}\hat{r}}=\bar{E}\text{.}
\label{ff}
\end{equation}

Then, if $\lim_{r\rightarrow 0}\rho =\rho _{1}$, it follows from (\ref{V})
and (\ref{m'}) that 
\begin{equation}
E\rightarrow \frac{8\pi \rho _{1}}{3}\text{, }\bar{E}\rightarrow \frac{8\pi
\rho _{1}}{3}\text{, }H\rightarrow \frac{8\pi \rho _{1}}{3}\text{,}
\end{equation}%
so the finiteness of $\rho _{1}$ entails that of the geometry.

It is worth noting that for the class of metrics under consideration $%
\lim_{\rho \rightarrow 0}m(r)=0$ according to (\ref{m}). Thus, our
configurations are such that in this limit we arrive not at the
Schwarzschild metric but at the Minkowski one.

Let 
\begin{equation}
f(\rho )=\chi (\rho )(\rho _{1}-\rho )  \label{f1}
\end{equation}%
with $\chi (\rho )>0$ everywhere finite including $\rho =\rho _{1}$. The
density $\rho =\rho _{1}$ at $r=0$ and it is seen from (\ref{ro}) that it is
decreasing monotonically. Near $r=0$, we get%
\begin{equation}
r\sim (\frac{\rho _{1}-\rho }{\rho _{1}})^{\frac{1}{2\chi _{1}}},
\end{equation}%
\begin{subequations}
\begin{equation}
\rho \approx \rho _{1}-Br^{2\chi _{1}}\text{,}  \label{rob}
\end{equation}%
where $\chi _{1}=\chi (\rho _{1})$ and $B>0$ is some constant.

\section{Compact configuration with a regular center}

By definition, a compact configuration borders with empty space in the outer
region $r>r_{0}$, so the metric is the ~Schwarzschild one there with the
mass $m_{0}$ and $\rho =0.$ The general condition of smooth joining two
pieces requires $p_{r}=0$ on the boundary $r=r_{0}$. As the equation of
state for the system under discussion is (\ref{eos}), this entails also

\end{subequations}
\begin{equation}
\rho (r_{0})=0.  \label{ro0}
\end{equation}%
It is worth mentioning that for smooth joining the equation $m(r_{0})=m_{0}$
is valid as well \cite{msh}, \cite{mass}. Assuming that the configuration
under discussion is a black hole, we identify $r_{0}$ with the horizon, $%
r_{0}=2m_{0}$. Although the horizon is a null surface, the continuity of the
metric function and its derivative is sufficient for smooth gluing. In this
aspect, we perform the same procedure that was exploited in refs. \cite%
{ov-k1} - \cite{ov-k2}.

Finally, choosing the constants accordingly, we have%
\begin{equation}
\frac{r}{r_{0}}=\exp (-\frac{1}{2}\int_{0}^{\rho }\frac{d\rho ^{\prime }}{%
f(\rho ^{\prime })}\equiv F(\rho )\text{.}  \label{int}
\end{equation}

This equation represents realization of the general formula (\ref{ro}) for
the case of a black hole that borders with an empty space due to the proper
boundary condition (\ref{ro0}) and the low limit of integral in (\ref{int}).
For regular configurations, the function $f(\rho )$ has the form (\ref{f1}).

For the mass we have for $r\leq r_{0}$%
\begin{equation}
m=4\pi r_{0}^{3}\int_{\rho }^{\rho _{1}}F^{\prime }(\bar{\rho})F^{2}(\bar{%
\rho})\bar{\rho}d\bar{\rho}.
\end{equation}

\subsection{Examples}

\subsubsection{Linear equation of state}

Let 
\begin{equation}
p_{\theta }=w\rho -(w+1)\rho _{1}\text{.}  \label{pw}
\end{equation}%
Then,

\begin{equation}
f=(w+1)(\rho -\rho _{1}).  \label{fw}
\end{equation}

The constants in (\ref{pw}) are adjusted to comply with (\ref{f1}). We also
require $w<-1$. In contrast to the standard phantom case, $p_{\theta }$ is
not proportional to $\rho $ but contains also the term with $\rho _{1}$. As
a result, $p_{\theta }+\rho >0$ for any $0<r\leq r_{0}$.

Now,%
\begin{equation}
\rho =\rho _{1}[1-\left( \frac{r}{r_{0}}\right) ^{2\left\vert w+1\right\vert
}]\text{.}
\end{equation}%
In contrast to (\ref{rob}), this is an exact relation valid for any $0\leq
r\leq r_{0}$. For the mass one has%
\begin{equation}
\frac{m}{m_{0}}=\frac{1}{2\left\vert w+1\right\vert }\left( \frac{r}{r_{0}}%
\right) ^{3}(2\left\vert w\right\vert +1-3\frac{r^{2\left\vert
w+1\right\vert }}{r_{0}^{2\left\vert w+1\right\vert }}).  \label{mm0}
\end{equation}

\subsubsection{Nonlinear equation of state}

In this manner, we can consider also nonlinear equations of state. Say, let
us take%
\begin{equation}
f=A(\rho _{1}-\rho )(\rho _{2}-\rho )
\end{equation}%
with $\rho <\rho \,_{2}<\rho _{1}$ and $A>0,$ where $\rho _{1}$, $\rho _{2}$
and $A$ are constants. Then%
\begin{equation}
\rho =\frac{\rho _{2}-z\rho _{1}}{1-z}>0\text{ }
\end{equation}%
since 
\begin{equation}
z\equiv (\frac{r}{r_{0}})^{2/\alpha }\frac{\rho _{2}}{\rho _{1}}\leq \frac{%
\rho _{2}}{\rho _{1}}<1\text{.}
\end{equation}%
Here,%
\begin{equation}
\alpha =\frac{1}{A(\rho _{1}-\rho _{2})}.
\end{equation}

It is convenient to rewrite $z$ in the form%
\begin{equation}
z=\left( \frac{r}{r_{1}}\right) ^{2/\alpha }\text{,}
\end{equation}%
where%
\begin{equation}
r_{1}=r_{0}\left( \frac{\rho _{2}}{\rho _{1}}\right) ^{\alpha /2}\text{.}
\end{equation}

According to (\ref{m}),%
\begin{equation}
\frac{m}{4\pi }=\rho _{1}\frac{r_{1}^{3}}{3}+(\rho _{2}-\rho _{1})r_{1}^{3}J(%
\frac{r}{r_{1}})\text{,}
\end{equation}%
where%
\begin{equation}
J(x)\equiv \int_{0}^{x}\frac{dx^{\prime }x^{\prime 2}}{1-x^{\prime 2/\alpha }%
}\text{, }  \label{j}
\end{equation}%
$x<1$.

In general, the integral (\ref{j}) cannot be reduced to elementary
functions. However, this can be done in some particular cases. Let, say, $%
\alpha =2/3$. Then,%
\begin{equation}
J(x)=-\frac{1}{3}\ln (1-x^{3})\text{.}
\end{equation}

\section{Dispersed systems}

We can include in our scheme systems without a sharp boundary. Such a case
was realized in \cite{dym}. We require $\rho \rightarrow 0$ when $%
r\rightarrow \infty $. Then, eq. (\ref{ro}) gives us%
\begin{equation}
r=r_{0}\exp [\frac{1}{2}\int_{\rho }^{\rho _{0}}\frac{d\rho ^{\prime }}{%
f(\rho ^{\prime })}]\text{,}  \label{dis}
\end{equation}%
where $\rho _{0}=\rho (r_{0})>0$ and $r_{0}$ are constants. The density is a
decreasing function of $r$, provided $f>0$. We assume that near $\rho =0\,\ $%
\begin{equation}
f\approx B\rho \text{,}  \label{fb}
\end{equation}%
where $B>0$ is a constant. Then, for $r\rightarrow \infty $%
\begin{equation}
\rho \sim r^{-2B}.
\end{equation}

The total mass is finite, if 
\begin{equation}
B>\frac{3}{2}.  \label{b}
\end{equation}

If eq. (\ref{f1}) is still valid near $r=0$, $\rho =\rho _{1}$, the center
is regular. Otherwise, it is singular.

\subsection{Examples}

\subsubsection{Linear equation of state and Kiselev's black hole}

Now, instead of eq. (\ref{f1}), let us take 
\begin{equation}
p_{\theta }=w\rho \text{, }f=(w+1)\rho   \label{lin}
\end{equation}%
with $w>-1$. Then, $f>0$ for any $\rho >0$. If we also admit the constant
term $m_{1}$ in the mass, we have from (\ref{V}), (\ref{m'}) and (\ref{dis})
that 
\begin{equation}
V=1-\frac{2m_{1}}{r}-(\frac{r_{1}}{r})^{2w}\text{,}
\end{equation}%
where $r_{1}$ is a new constant. This corresponds precisely to eq. (14) of 
\cite{kis}, if we redefine $w=\frac{1+w_{q}}{2}$, where notation $w_{q}$ was
used in \cite{kis}. If $w>0$, matter extends to infinity where $\rho
\rightarrow 0$. The positivity of density requires $w<\frac{1}{2}$. 

If $w=0$,%
\begin{equation}
\rho =\frac{const}{r^{2}}.  \label{-2}
\end{equation}

If $m_{1}=0$ and $w\leq -1$, the metric is regular near the center according
to (\ref{e}) - (\ref{ff}). But in this case there exists a cosmological
horizon at $r=r_{1}$.

\subsubsection{Nonlinear equation of state}

\begin{equation}
f=(1+w)(\rho -\frac{\rho ^{2}}{\rho _{1}})
\end{equation}%
with $w>-1$. Now,%
\begin{equation}
\rho =\frac{\rho _{1}}{1+z}\text{, }z=\left( \frac{r}{r_{1}}\right)
^{2(w+1)}.
\end{equation}%
\begin{equation}
\frac{m}{4\pi }=\rho _{1}r_{1}^{3}I(\frac{r}{r_{1}})\text{,}
\end{equation}%
where 
\begin{equation}
I(x)=\int_{0}^{x}dx^{\prime }\frac{x^{\prime 2}}{1+x^{\prime 2(w+1)}}\text{.}
\end{equation}%
When $r\rightarrow 0$, $\rho \rightarrow \rho _{1}$ and when $r\rightarrow
\infty $, $\rho \rightarrow 0$. If, according to (\ref{b}), $w>\frac{1}{2}$,
the total mass is finite.

In particular, for $w=0$ and $r\rightarrow \infty $ we get again%
\begin{equation}
\rho \sim r^{-2}\text{.}  \label{r2}
\end{equation}%
It is worth noting that dependence (\ref{-2}), (\ref{r2}) appears also in
another context connected with a strong gravitational mass defect \cite{zel}.

In doing so,%
\begin{equation}
I(x)=x-\arctan x\text{.}
\end{equation}

\subsubsection{Dymnikova's black hole}

Let us choose 
\begin{equation}
f(\rho )=\frac{3}{2}\rho \ln \frac{\rho _{1}}{\rho }\text{, }p_{\theta
}=-\rho +f(\rho )\text{,}
\end{equation}%
where $\rho _{1}=\rho (0)>\rho _{0}=\rho (r_{0})$.

Then, calculating the integral in (\ref{dis}), one obtains%
\begin{equation}
\rho =\rho _{1}\exp (-A\frac{r^{3}}{r_{0}^{3}})\text{,}
\end{equation}%
where $A=\ln \frac{\rho _{1}}{\rho _{0}}$. This coincides with eq. (8) of 
\cite{dym} in somewhat different notations.

\section{Conclusions}

We have thus described the whole class of solutions. We do not need to
invent the dependence $m(r)$. Instead, we rely on a more physical entity -
the equation of state. Following this line, we managed to find in closed
form the relation between the radius and the energy density. The solution is
somewhat unusual in what we found formally the dependence $r(\rho )$ instead
of the usual dependence $\rho (r)$. The corresponding formulas are valid for
any equation of state, including nonlinear ones. For a wide class of
physically reasonable equations of state $p_{\theta }=p_{\theta }(\rho )$
the formulas can be inverted to give $\rho (r)$. Our approach applies to
both compact and dispersed systems.

It is of interest to extend our approach to rotating systems, charged black
holes, and cosmological solutions.

\section{Data availability}

No data was used for the research described in the article

\section{Declaration of competing interest}

The author declares that he has no known competing financial interests or
personal relationships that could have appeared to influence the work
reported in this paper.

\section{Acknowledgement}

I thank Te\'{o}filo Vargas and Grupo de Astronom\'{\i}a SPACE and Grupo de F%
\'{\i}sica Te\'{o}rica GFT, Facultad de Ciencias F\'{\i}sicas, Universidad
Nacional Mayor de San Marcos, Lima-Per\'{u} for hospitality. This work was
supported in part by the grants 2024/22940-0, 2021/10128-0 of S\~{a}o Paulo
Research Foundation (FAPESP).


\begin{thebibliography}{99}
\bibitem{ans} S. Ansoldi, Spherical black holes with regular center,
[arXiv:0802.0330].

\bibitem{glin} E. B. Gliner, Algebraic properties of the energy-momentum
tensor and vacuum-like states of matter, Sov. Phys. JETP 22 (1966) 378.

\bibitem{dym} I. G. Dymnikova, Vacuum nonsingular black hole, General
Relativity and Gravitation 24 (1992) 235.

\bibitem{bron} K. A. Bronnikov, Regular black holes as an alternative to
black bounce, Phys. Rev. D 110 (2024) 024021 [arXiv:2404.14816].

\bibitem{ov1} J. Ovalle, Schwarzschild black hole revisited: Before the
complete collapse, Phys. Rev. D 109 (2024) 104032, [arXiv:2405.06731].

\bibitem{ov-k1} R. Casadio, J. Ovalle, A. Kamenshchik, Cosmology from
Schwarzschild black hole revisited, Phys. Rev. D 110 (2024) 044001
[arXiv:2407.14130].

\bibitem{ov-k2} R. Casadio, J. Ovalle, A. Kamenshchik, Regular Schwarzschild
black holes and cosmological models, Phys. Rev. D 111 (2025) 064036
[arXiv:2502.13627].

\bibitem{bard} J. M. Bardeen, Non-singular general-relativistic
gravitational collapse, in Proceedings of International Conference GR5
(Tbilisi, USSR, 1968), p. 174.

\bibitem{vis} A. Simpson and M. Visser, Black bounce to traversable
wormhole, JCAP 02 042 (2019), [arXiv:1812.07114].

\bibitem{kis} V. V. Kiselev, Quintessence and black holes, Classical and
Quantum Gravity 20 (2003) 1187, [gr-qc/0210040].

\bibitem{sant} L. C. N. Santos, Regular black holes from Kiselev anisotropic
fluid, Eur. Journ. of Phys. C, 84 (2025) 1318, [arXiv:2411.18804].

\bibitem{mtw} C.W. Misner, K. S. Thorne, and J. A. Wheeler, Gravitation
(Freeman, San Francisco, 1973).

\bibitem{msh} C. W. Misner and D. H. Sharp, Relativistic Equations for
Adiabatic, Spherically Symmetric Gravitational Collapse, Phys. Rev. 136,
8571 (1964).

\bibitem{mass} M. E. Cahill and G. C. McVittie, Spherical Symmetry and
MassEnergy in General Relativity. I. General Theory, J. of Math. Phys. 11,
1382 (1970).

\bibitem{zel} Y. B. Zel'dovich, The collapse of a small mass in the general
theory of relativity, Zh. Eksp. Teor. Fiz. 42, 641 (1962) [Sov. Phys. JETP
15, 446 (1962)].
\end{thebibliography}
\end{document}